\documentclass[conference]{IEEEtran}
\IEEEoverridecommandlockouts
\usepackage{cite}
\usepackage{amsmath,amssymb,amsfonts}
\usepackage{algorithmic}
\usepackage{graphicx}
\usepackage{textcomp}
\usepackage{xcolor}
\def\BibTeX{{\rm B\kern-.05em{\sc i\kern-.025em b}\kern-.08em
    T\kern-.1667em\lower.7ex\hbox{E}\kern-.125emX}}
\begin{document}

\title{Campus Wi-Fi Coverage Mapping and Analysis}

\author{\IEEEauthorblockN{\small Farhana Binte Kamrul Easha}
\IEEEauthorblockA{\textit{\small School of Engineering (Telecommunications)} \\
\textit{\small Macquarie University}\\
\small New South Wales, Australia \\
\small farhana-binte-kamrul.easha@students.mq.edu.au}
\and
\IEEEauthorblockN{\small Robert Abbas}
\IEEEauthorblockA{\textit{\small School of Engineering (Telecommunications)} \\
\textit{\small Macquarie University}\\
\small New South Wales, Australia \\
\small robert.abbas@mq.edu.au}
\and
\IEEEauthorblockN{\small Matthew Daley}
\IEEEauthorblockA{\textit{\small IT- Infrastructure and Applications} \\
\textit{\small Macquarie University}\\
\small New South Wales, Australia \\
\small matthew.daley@mq.edu.au}
}

\maketitle

\begin{abstract}
Wireless Local Area Networks (WLANs), known as Wi-Fi, have become an essential service in university environments that helps staff, students and guests to access connectivity to the Internet from their mobile devices. Apart from the Internet being a learning resource, students also submit their assignments online using web portals. Most campuses will have poor coverage areas for mobile networks and, as a result, the ability of the wireless network to supplement Internet access for mobile devices in these areas becomes more important. Acquiring clear understanding of WLAN traffic patterns, network handover between access points and inter-network handover between the Wi-Fi and mobile networks, the optimal placement of networking equipment will help deliver a better wireless service. This paper presents data analyses and Wi-Fi signal coverage maps obtained by performing wireless radio surveys, coverage predictions and statistical analysis of data from the existing access points to show the current Wi-Fi performance in several locations of a large university campus. It them makes recommendations that should improve performance. These recommendations are derived from AP performance testing and made in the context of cabling length limitations and physical and aesthetic placement restrictions that are present at each location.
\end{abstract}


\section{Introduction}

The Internet has grown at a remarkable pace since the roots of its inception with APRANET in the 1960s. It has been a large driver of many business efficiencies for the last thirty years. Handwritten letters sent through post have been surpassed by texts and emails, long queues at the bank counters has turned into in application lines for visit support, downloading eBooks onto a Kindle or iPad has overtaken the physical trips to libraries and individuals have the option to make their own customized websites to mirror their interests. As an ever-increasing number of people came online, the enthusiasm for simpler access expanded. Wi-Fi became an integral part of enabling this access. Accessing the Internet, through Wi-Fi, became easier because laptops and other mobile devices could simply be brought to a nearby Wi-Fi access point to gain access.

Wi-Fi provides Internet access in private homes and organizations, as well as in open commercial spaces offering Wi-Fi hotspots set up either complimentary or commercially. Organizations, for example, shopping centres, hotels or airports, regularly offer free hotspots to attract customers. According to a statistical data on global public Wi-Fi hotspots \cite{b1}, projections suggest that by the end of 2020 there will be an aggregate of 454 million open Wi-Fi hotspots around the world when compared to 94 million in 2016. In schools, universities and colleges, Internet access has been available for many years and its usage is transforming beyond just a helpful resource or a piece of IT technology. From online webinars and live classes, to digital books and other online assets, the internet and has turned learning into an incredible teaching tool in education. The first campus wide wireless internet network was constructed in Carnegie Mellon University (CMU), at its Pittsburgh campus in 1993 which was named as “Wireless Andrew”. The CMU Wi-Fi zone was completely operational by February 1997 \cite{b2}. Conventional academic campuses in the present day provide at least partial campus Wi-Fi coverage as both the teaching staff and students require connectivity to the internet for accessing university portals, availing study materials and online resources. 

There are numerous specifications of IEEE 802.11 (Wi-Fi), for example, 802.11a, 802.11b, 802.11g, 802.11n and 802.11ac.  The specifications named as 802.11n and 802.11ac (also known as Wi-Fi 4 and Wi-Fi 5 respectively) are the most widely used at present. Wi-Fi is expected to get faster and better with its latest update IEEE 802.11ax (also known as Wi-Fi 6). While many Wi-Fi routers now exist with capabilities utilizing the draft IEEE 802.11ax specification, it wasn’t truly finalised until the Wi-Fi 6 certification became official in September 2019. This finalisation introduced a flood of updated devices to the market advertising new wireless capabilities offering towards much faster and less congested next generation networks. The new standard (IEEE 802.11ax) is the successor to 802.11ac (Wi-Fi 5) with numerous updated features, which provides significantly faster maximum speeds (4x to 10x faster than the previous standards) by combining the 2.4 GHz and 5 GHz spectrum bands. MU-MIMO (4x4 MIMO) technology increases the total bandwidth capacity to be as high as 14 Gbps. This latest version of Wi-Fi ought to be the fundamental use from 2020 and may also be called “High-efficiency Wireless”  \cite{b3}. Wi-Fi 6 is likely to be implemented more quickly than past Wi-Fi standards, as it will show an immediate improvement on network performance in crowded locations like university campuses, apartments and stadiums. 

In this study, we focus on a large university campus, the kind which is commonly present in most of the cities around the world. After analysing the network traffic data from the campus wireless network, we perform the following tasks: (i) We assess the performance of the existing wireless network; (ii) we identify the locations of most congested access points and (iii) we identify problem areas with poor or no signal coverage throughout the campus. To do this, we first generate wireless signal coverage maps in selected locations using wireless measurement tools and, separately, study network data collected from the Wi-Fi management platform. We then identify the busiest campus access points by looking at the number of connected clients to each access point and list the problem areas having wireless signal strength level below a reference level. Finally, we propose suitable solutions to the problems and present coverage predictions of the wireless network if the proposed solution were implemented.

This paper is organized as follows: Section II reviews the relevant literature on campus Wi-Fi; Section III explains the reference scenario along with the techniques and tools we used throughout the experiments. The performance analysis of the existing Wi-Fi network and traffic flow characteristics of the access point are presented in Section IV; Section V contains few plans to improve the campus wireless network. Section VI concludes the work along with directions for future work.

\section{Literature Review}

The assessment of data logs obtained from wireless networks has been playing a vital role in the way of achieving the optimal network performance. Researchers utilize different techniques to perform experimentation and different analyzing methods.

In their work, \cite{b4}, Redondi, Cesana, Weibel and Fitzgerald investigate the use of a wireless network deployed in a large-scale technical university. They leverage three-weeks of Wi-Fi traffic statistics and characterize the spatio-temporal correlation of the traffic at different granularities (individual access point, groups of access points, entire network).

Tang and Baker in \cite{b5}, analyze a twelve-week trace of a campus wireless network which contains overall user behavior, for example, when and how intensively people use the network and how much they move around, overall network traffic and load characteristics by observing throughput and symmetry of incoming and outgoing traffic, and traffic characteristics from a user point of view. 

In \cite{b6}, the information was collected from 12 indoor access points belonging to a university wireless network with a view to verify loads on the access points and evaluate performance of the local university wireless network. the data were obtained every five minutes through passive experiment for a 10-weeks period. 

The main focus of Kotz and Essien in \cite{b7}, is the evaluation of a campus Wi-Fi network consisting 476 access points. Traffic and association related information are collected for a two-month period through Simple Network Management Protocol (SNMP) polling and System Logging Protocol (syslog) messaging. A further passive data collection compliments the initial active data gathering in order to capture traffic at back-end. They aim on analyzing the  dataflow type traits (dataflow summary for each application), dataflow load features (per client, per access point, per building, dataflow variability over time and user mobility. After two years, Henderson, Kotz and Abyzov completed a related analysis over a seventeen-week span on the same Wi-Fi network in \cite{b8}, to assess the changes of the above-mentioned performance statistics over time. The measurement techniques employed include SNMP polling, telephone records, syslog and tcpdump packet sniffing. The new trace indicates major upturns in streaming multimedia, peer-to-peer and voice over IP (VoIP) dataflow; whereas the initial WLAN usage was dominated by web dataflow.

Sulaiman and Yaakub in \cite{b9}, work in 225 locations to investigate the performance analysis, network auditing and connectivity problems of Wi-Fi network in a campus network. An unreliable transfer of data and high data corruption was detected due to overlapping operational frequencies as many access points utilize the same channel. 

In \cite{b10}, a comparative study of traffic flow from handheld and non-handheld devices is presented. They analyze the traffic flow characteristics from two different campus wireless networks for a period of 3 days for 32,278 unique devices.

As it is still needed to update the research on performance analysis of campus wireless networks and characterize new problems solutions, our work significantly focuses on network traffic statistics, coverage statistics and performance analysis of a mature campus wireless network. Furthermore, solutions to various problems that cause the existing network to fall back from providing optimal wireless service has been addressed. 

\section{Scenario}
In this section we illustrate the environment of our analysis, the campus of Macquarie University along with the methods and tools used to take the study ahead.  

\subsection{Existing Network Structure}
This work explores the network data obtained from the wireless network that encompasses the entire university campus spanning over 126 hectares. This analyzed wireless network is classified as a high density (HD) network consisting 2000 access points that provides Wi-Fi facility to various indoor and outdoor environments, for example, academic and administrative buildings housing the lecture rooms, laboratories, offices and conference spaces, library, theatre halls, food court and café, fitness center and other public spaces. All access points share the same SSID that offers a smooth Wi-Fi experience to the roaming wireless users affiliated with the university. The Management layer of the WLAN comprises of the Management Platform which basically offers a sole point for managing the network. It includes centralized configuration, report presenting, coverage maps, investigating and troubleshooting. There is a variety of access points installed around the university campus to suit specific environment. These access points support the IEEE standards 802.11n, 802.11ac and 802.11ax in combination. Indoor access points are commonly conveyed in one of two designs: ceiling mounted, or wall mounted. Cubicle and workspace mounts are rarely employed as normally a clear line of sight is not taken into consideration by these settings, which thus can cause reduced wireless network performance. Most of current WLAN arrangements are at the ceiling level. A ceiling arrangement can be made at or underneath the ceiling level. Wall mounted access point setup is not as common as ceiling mounted deployments, yet may be found in huge spaces, for example, auditoriums or lecture halls where the ceilings are difficult to reach or in case of fixed ceilings where wiring is a big challenge. Similarly, wireless access points are implemented in various outdoor locations around the university campus in order to offer an uninterrupted wireless experience all through the grounds. Basically, an outdoor wireless access point directly associates to a wired Ethernet connection and subsequently delivers wireless connections to various devices utilizing radio frequency links. 
In this work, we observed the network traffic statistics on the Management Platform and the coverage statistics for the time period February 25, 2019 to July 09, 2019 (duration of Session-1 for the year 2019) with a view to identify the network black spots as well as the congested areas around the campus where existing wireless service requires to be improved. Initially, we focused on identifying the busiest day of the week as well as the busiest time of the day according to the number of clients connected to various access points throughout the campus. Following the data pattern, we could notice that the much busier days of a week lie between Tuesdays to Thursdays; whereas Fridays to Mondays seem to be less busy, Sundays being the least as depicted in figure 1. Next, on a daily basis, the number of users connected to the network increases gradually around 7 o’clock in the morning as staffs and students pass in the campus, and between 12:15 and 4:30 in the afternoon seems to be the peak hours with the highest number of connected clients. As the day comes to an end, the number of connected clients starts reducing drastically around 6 o’clock in the evening as most of the users start leaving campus by this time and lastly the least number of users shows up after 11 o’clock at night (students and visiting academics living in campus). The data usage (from clients and to the clients) shows a similar behavior. The number of connected client and the data usage over a period of 24 hours is shown in figure 2. 

\begin{figure}[htp]
\centerline{\includegraphics[width=8cm]{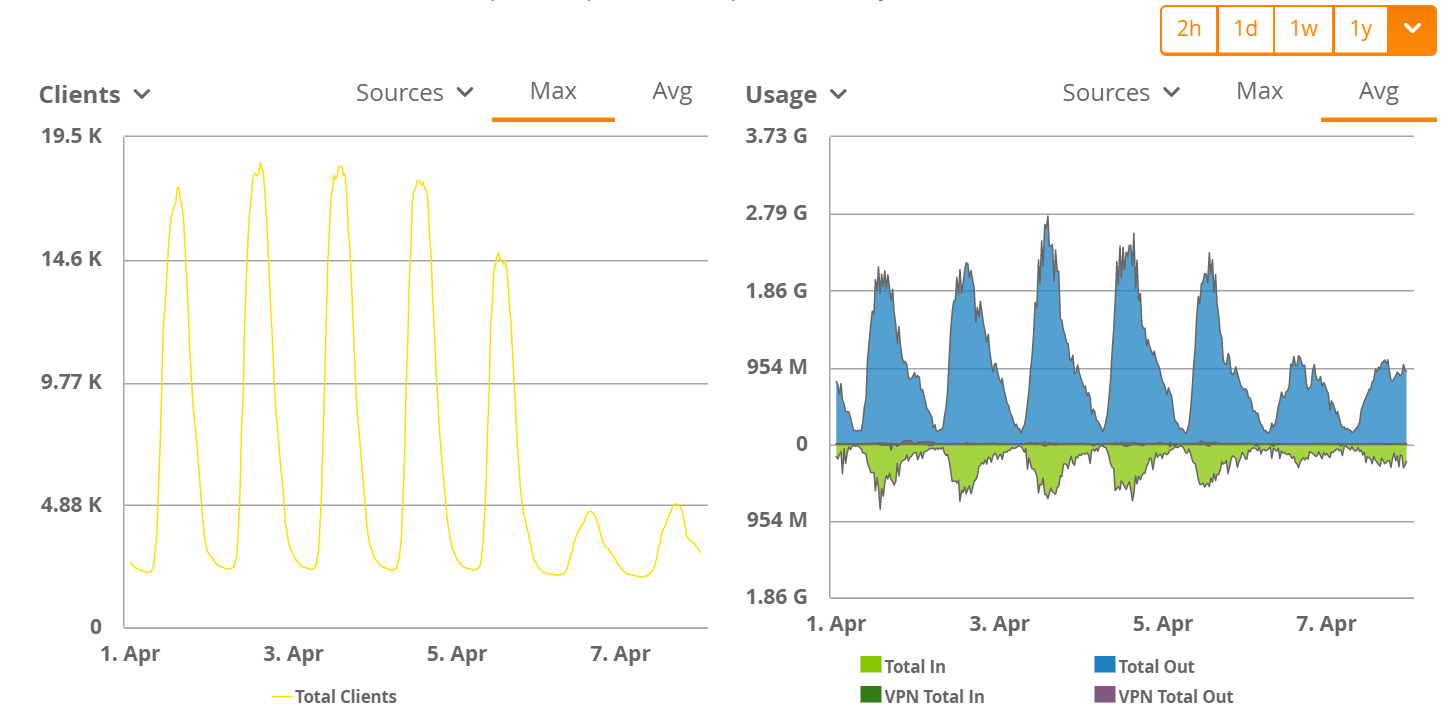}}
\caption{Number of connected clients and data usage over a week}
\label{fig1}
\end{figure}

\begin{figure}[htp]
\centerline{\includegraphics[width=8cm]{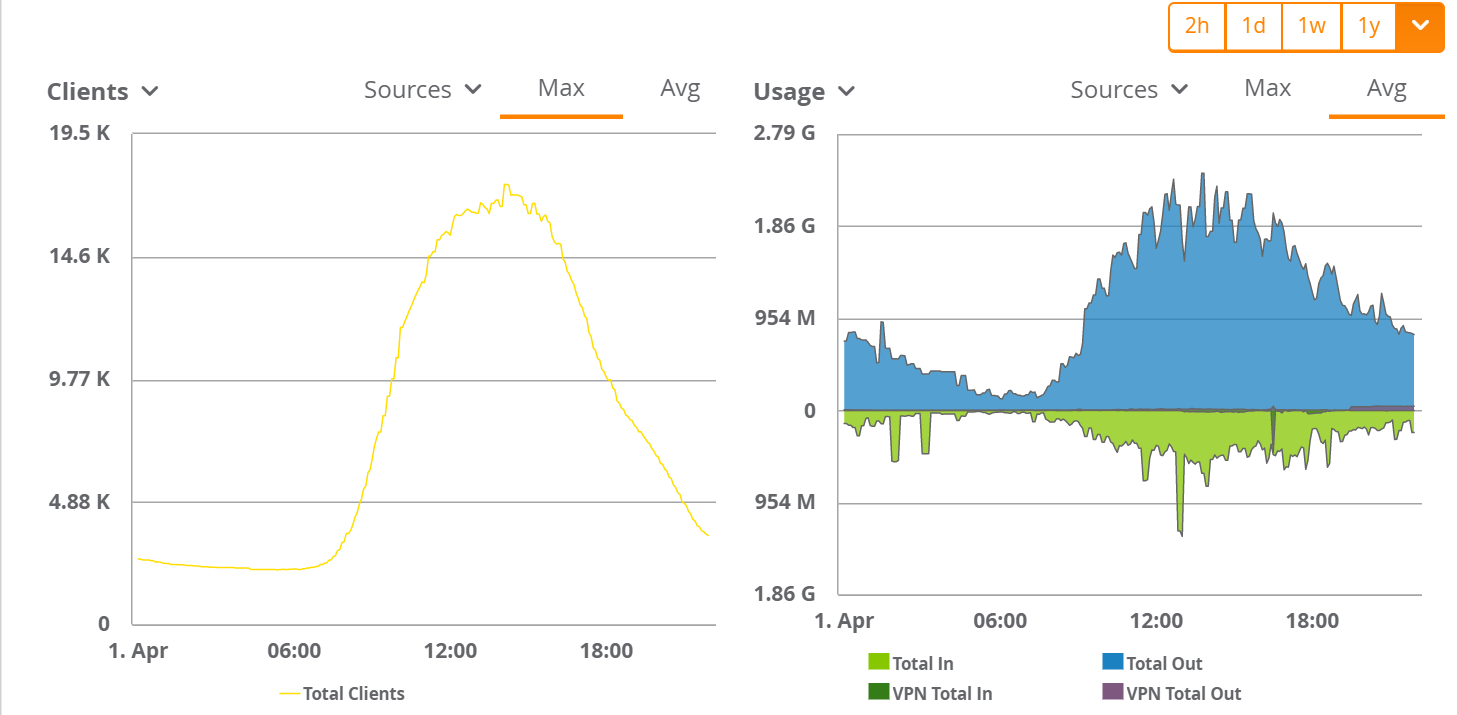}}
\caption{Number of connected clients and data usage over a day}
\label{fig2}
\end{figure}

\subsection{Methodology and Components Used}
Deployment of a wireless network is much complex compared to a wired network. In contrast to wired networks, over-provisioning of access points is definitely not an excellent thought with wireless systems, as the possibility of co-channel interference gets high with many access points located adjacent. Therefore, wireless measurements are significant with an ultimate objective to boost wireless signal quality as well as reduce bad indoor and outdoor signal propagation. This significance of wireless measurements encouraged many wireless device manufacturers to introduce supervision for the same. Cisco being one of them, has specified a wide-ranging guidance for executing wireless measurements in Wi-Fi deployment. Ekahau, a wireless measurements tool vendor has a partnership with Cisco. There are numerous other wireless measurements tools available, for example NetSpot, NetScout, AirMagnet Survey, VisiWave, Wi-Fi Scanner, TamoGraph etc. These wireless measurements tools support IEEE 802.11 a/b/g/n/ac/ax wireless systems and can be run on both or any of Microsoft Windows and macOS. One can discover the parameters, for example, signal quality, Signal-to-Noise Ratio (SNR), operating frequency and channels, location of access points, noise floor and so forth as well as conduct a wireless measurements walking through the site grid having any of these measurements tools installed in his hand-held device. 
In this work, initially we performed several wireless measurements using Ekahau measurements tool installed on our movable devices; but as its performance was not completely satisfactory, we started using NetSpot which offers more extensive data analyzing scopes. Among the two types of wireless measurements, i.e. actual measurements and predictive measurements, we opt for the actual (active) wireless measurements depending on the studied environment. In this method, we obtained the floor-plans for several predetermined locations around the campus and obtained the wireless signal coverage maps by collecting signal strength measurements on numerous sample points while walking through the grids. On the other hand, when it came to various data collection regarding the campus Wi-Fi network, we have been receiving a great support from the Management Platform. The full-featured Management Platform is scalable for mixed-vendor wired and wireless systems which provided us with a huge choice of information that includes total number of access points (both active and inactive) installed in both indoor and outdoor locations along with their frequency bands and associated channels, number of clients connected to these access points during various time periods, data usage by connected users (both uplink and downlink) and so on.

\section{Findings and Discussion}
In this section, we present the outcomes obtained from various wireless measurements performed in different locations on campus along with the analyzed network data.

\subsection{Wi-Fi Signal Strength Variation with Distance}

To get a more practical overview of the variation of Wi-Fi signal strength offered by an outdoor wireless access point over distance, we activated a temporary SSID “Paolo” (device model: AP-375, Aruba Networks) in a predetermined location near the food court (marked with black inked box in figure 3) and performed a wireless measurements, taking the signal strength measurements in sixteen different sample points (line of sight) by placing each of them 4.20 meters away from the previous one. The generated coverage map along with the obtained data is presented in figure 3 and figure 4 respectively.

\begin{figure}[htp]
\centerline{\includegraphics[width=8cm]{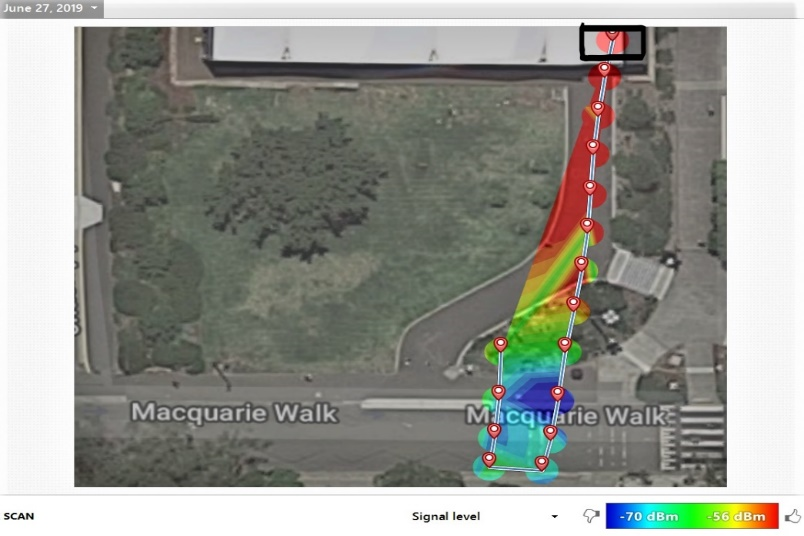}}
\caption{coverage map generated near food-court}
\label{fig3}
\end{figure}

\begin{figure}[htp]
\centerline{\includegraphics[width=8cm]{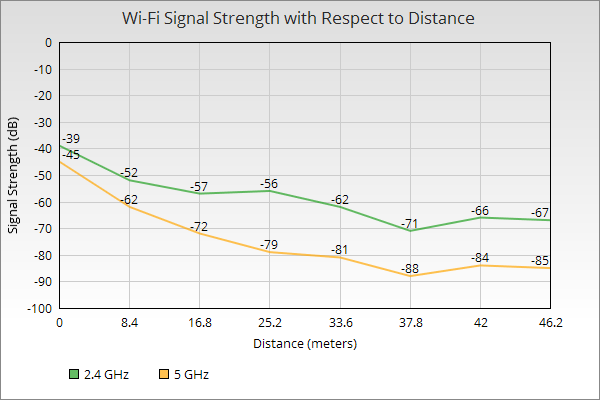}}
\caption{Wi-Fi signal strength over distance}
\label{fig4}
\end{figure}

From figure 3 and figure 4, its visible how both 2.4 GHz and 5 GHz wireless signals lose their strength with the increase in distance from the wireless access point. So, the maximum viable distance at which the minimum desired signal strength (-70 dBm) can be achieved is roughly 37 meters for the signals with frequency of 2.4 GHz, whereas the same seems to be only around 16 meters for the wireless signals with a higher frequency of 5 GHz. It is also noticeable that in few points the signal strength is reduced (blue in color), yet again it increases by some extend in other points although the signals are travelling farther away. This does not completely match with the theoretical explanation, but practically we may mention the plants and uneven heights of sample points as reasons for this matter.

\subsection{Practical Performance Analysis Based on Wireless Measurements}
This section represents the real-life scenario of Wi-Fi signal strength provided by indoor and outdoor access points. 

\paragraph{Performance Analysis in Indoor Environment}
 Figure 5 shows the coverage map generated by performing a NetSpot wireless measurements in the University Graduation Hall (E7B). For completing the survey, we utilized a temporary access point which was named as “Paolo” (device model: APIN0325, Aruba Networks) and was mounted vertically about 4 meters above the ground level (marked as a square box in figure 5). After turning on the access point, we selected the SSID- Paolo (5 GHz) having BSSID/MAC address A8:BD:27:CA:51:20, so that we could obtain the actual coverage map of the area solely covered by this specific access point. As shown in the figure, the signal strength is not equally distributed over the coverage area. Here, it's mentionable that the coverage map could have a diverse appearance if the device could be mounted horizontally at the ceiling level. 

\begin{figure}[htp]
\centerline{\includegraphics[width=8cm]{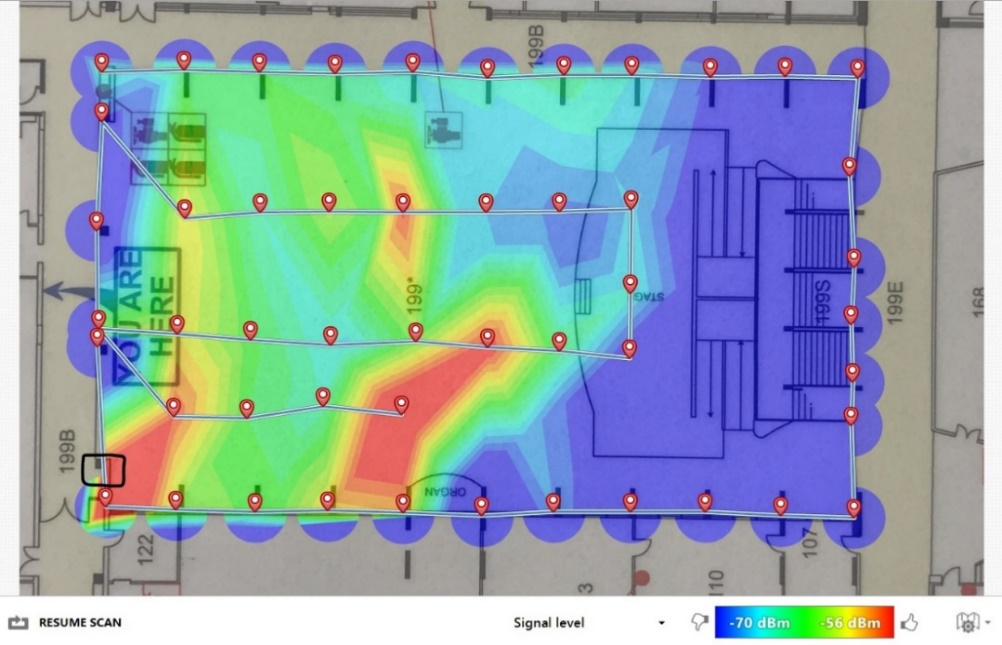}}
\caption{coverage map generated in Macquarie university graduation hall}
\label{fig5}
\end{figure}

\paragraph{Performance Analysis in Outdoor Environment}
Macquarie University Library has been found to be one of the busiest areas with a huge number of users connected to the Wi-Fi network throughout the day. As the location in front of the library is a conjunction area used by a huge population whether going to or from library, parking or any other destination, demand of a strong and uninterrupted Wi-Fi coverage is obvious. So, with the aim to find out the current Wi-Fi signal strength in the subject area, we initiated a NetSpot wireless measurements and the obtained coverage map is depicted in Figure 6. For completing the wireless measurements, Wi-Fi signal strength of campus network was measured in 24 sample points within the area surrounded by the Library, Central Avenue and Macquarie Walk. The sample points were chosen putting a gap of 8 meters from the previous ones. As we obtained the coverage map, the outcome was quite disappointing exhibiting a deep blue color over the whole area which represents a weak signal level, generally below -70 dBm. This is why, if someone just exits the library or walks through the area while talking over an internet call, experiences a call drop or needs to switch on the mobile data to continue. 

\begin{figure}[htp]
\centerline{\includegraphics[width=8cm]{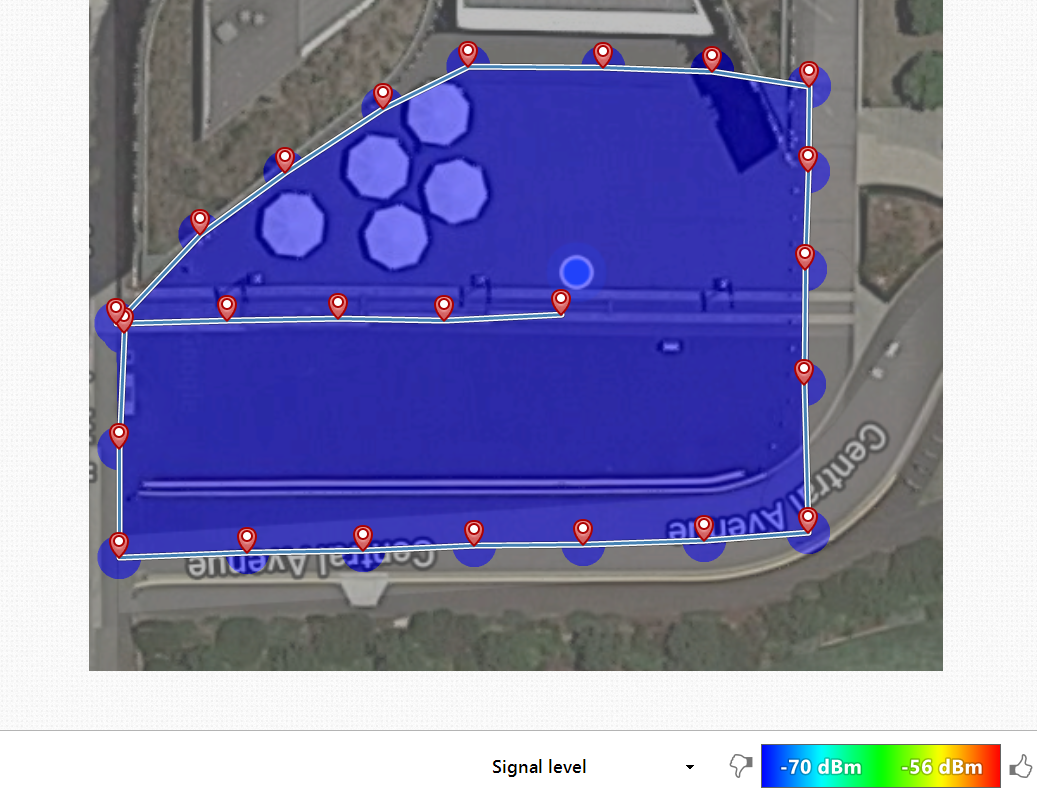}}
\caption{coverage map generated in front of Macquarie university library}
\label{fig6}
\end{figure}

\subsection{Practical Performance Analysis Based on Collected Data}

While collecting the data from the Management Platform and observing the same, we got an indication that Macquarie University Library situated in 16 Macquarie Walk (C3C) has the busiest access points installed inside it which portrays that students have frequent visits to the library before, after or may be during the break between their classes. The access points for which we observed and collected the data belong to two different levels of the library, level-3 and level-4 respectively. 

At level-3, altogether there are thirteen (13) active access points serving the Wi-Fi users. Among them, the access point named as C3C-3-AP8 is found to be most busy (normally in total above 40 devices connected to its either frequency bands). A group of data collected for the access point (C3C-3-AP8) on a monthly basis over the examined period is presented in figure 7. 

\begin{figure}[htp]
\centerline{\includegraphics[width=8cm]{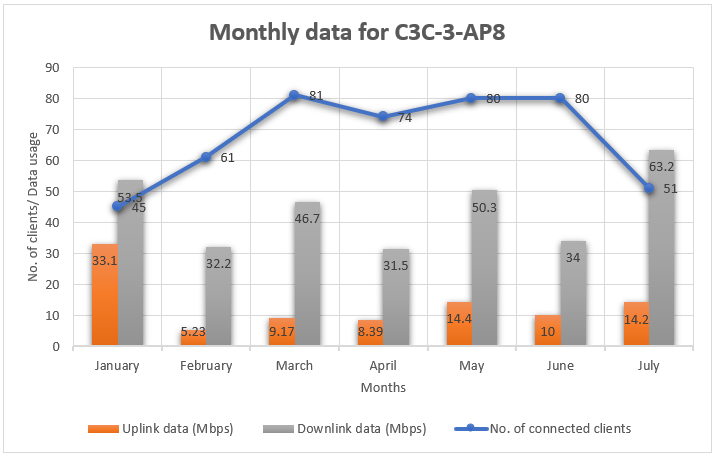}}
\caption{Maximum number of connected clients and data usage for C3C-3-AP8}
\label{fig7}
\end{figure}

Next, at level-4, currently there are eighteen (18) active access points. Among them, the access point named as C3C-4-AP5 is found to be the busiest one (normally altogether more than 35 users connected to any one of the two frequency bands). A collection of data for the access point (C3C-4-AP5) on a monthly basis over the studied period is presented in figure 8.

\begin{figure}[htp]
\centerline{\includegraphics[width=8cm]{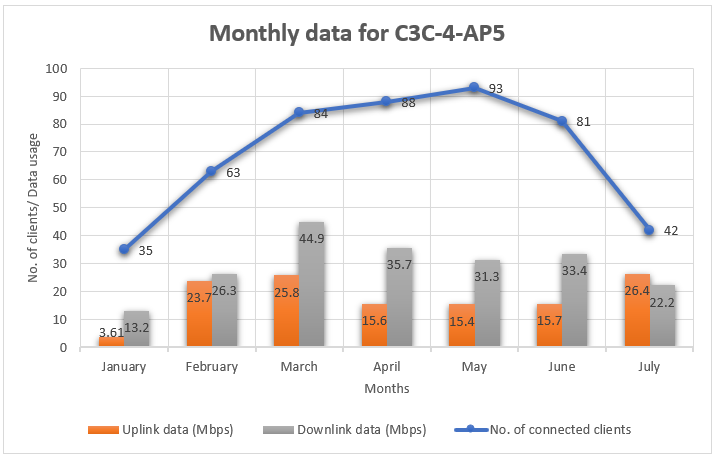}}
\caption{Maximum number of connected clients and data usage for C3C-4-AP5}
\label{fig8}
\end{figure}

\section{New Access Points Deployment Plans}
This section presents few plans for implementing new access points along with related calculations, execution of which will improve the overall performance of the existing wireless network.

\subsection{New AP Installation Plans Based on wireless measurements}

With a view to improve the Wi-Fi coverage near the library entrance, a ‘new access points installation plan’ has been proposed. Two new Omni-directional PoE (Power over Ethernet) Wi-Fi access points can be installed in two corners of the drop ceiling at library entrance (marked as AP-1 and AP-2 in figure 9).

Ethernet cables (Cat 5e/Cat 6) can be used here for data transmission. As the Ethernet cable has the limitation of transmitting data over a maximum distance of 100 meters, a precise calculation of the distances between the proposed access points installation locations and the nearest data communications room is required. Walking through the library, the nearest data communications room we could find is room number 217 in Library-level 2. Next, the distances from the new access points installation locations are calculated as follows

\textit{Distance between Communications Room and AP-1 (in meters):}

AP-1 to library L-2 entrance = 20.5 metes
Entrance to room no. 217 = 29.6 meters
Hence, total distance = 50 meters (approximately) 

\textit{Distance between Communications Room and AP-2 (in meters):}

AP-2 to library L-2 entrance = 31.2 metes
Entrance to room no. 217 = 29.6 meters
Hence, total distance = 61 meters (approximately) 

\begin{figure}[htp]
\centerline{\includegraphics[width=8cm]{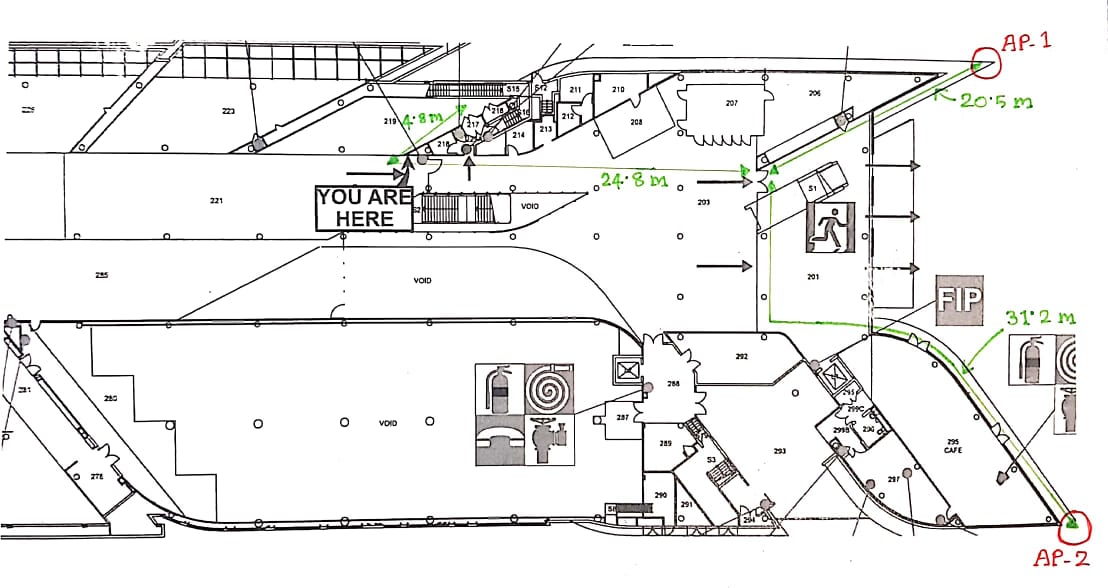}}
\caption{Measurements for required Ethernet cable for installation of new APs}
\label{fig9}
\end{figure}

As per the calculation shown above, the distances between the nearest communications room and both the proposed new AP installation locations are below 100 meters which means Ethernet cable can serve the purpose. The reason behind choosing PoE access points is that, installing new power lines at those locations will be inconvenient. Moreover, it is a cost-effective option to use a single cable for both data transmission and electric power. A graphical overview for the coverage prediction of improved Wi-Fi signal in the same area after installing the new access points is depicted in figure 10. The ceiling mounted donut shaped access points with built-in omni-directional antenna will provide a 360° wireless coverage with both vertical as well as horizontal connectivity. The Wi-Fi signal will have the strongest strength near the two access points, at areas marked as red in the figure and eventually will lose strength while travelling farther away at areas colored as blue. This may follow a pattern similar to the graph presented in figure 4.

\begin{figure}[htp]
\centerline{\includegraphics[width=8cm]{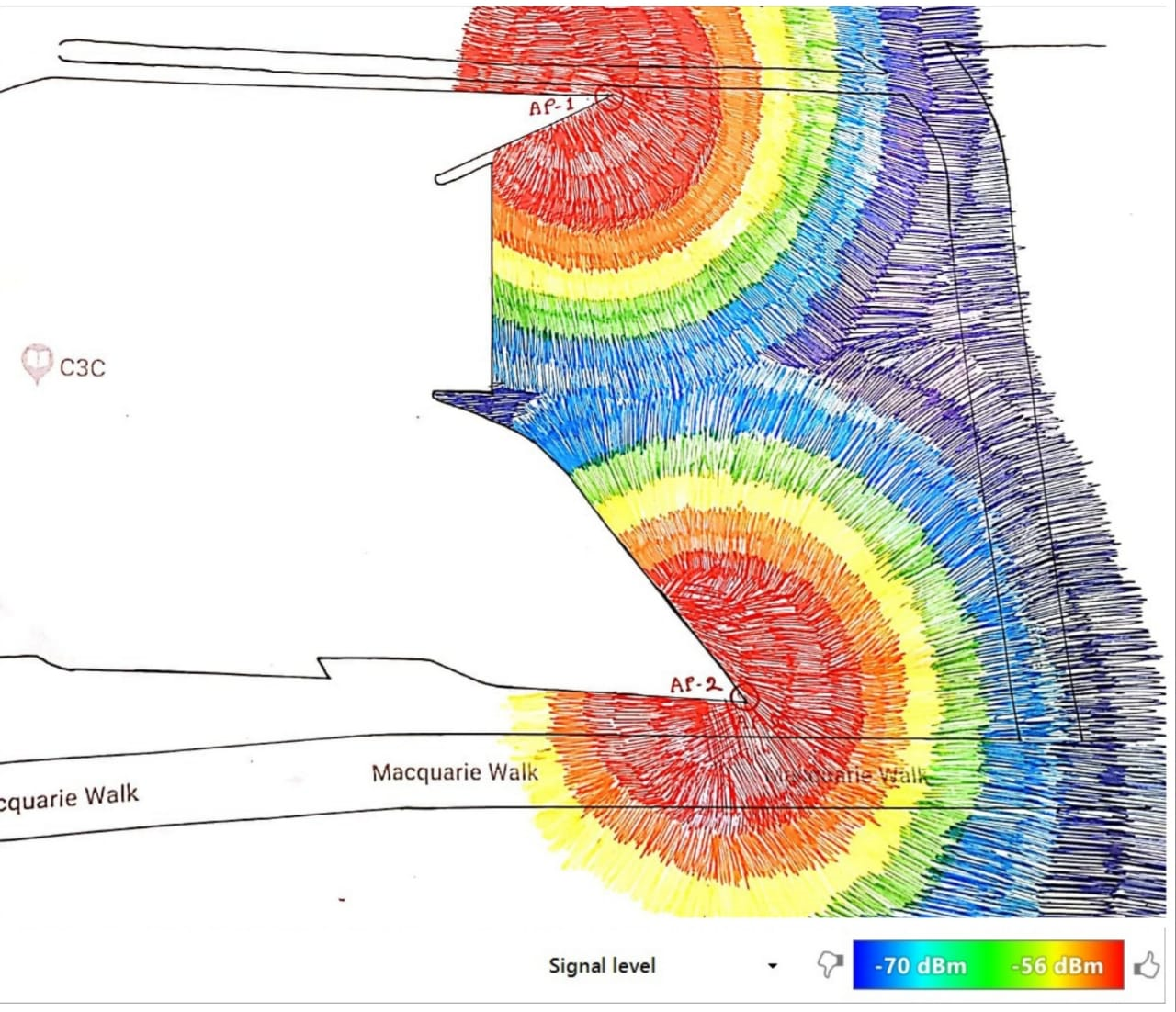}}
\caption{Wi-Fi coverage prediction post-deployment of new APs}
\label{fig10}
\end{figure}

\subsection{New AP Installation Plans Based on Data Analysis}

Following the data trend for library L-3, we came up with a suggestion to place a new access point near the examined access point (C3C-3-AP8) and the new floor plan is shown in figure 11 with the new AP marked as C3C-3-APX. To activate this new access point in the proposed location we will require approximately seven (07) meters of Ethernet cable to cover the distance between the new access point and the nearest communications room (room no. 342) and the newly installed access point will definitely reduce the excess pressure on C3C-3-AP8 by providing Wi-Fi connection to a reasonable number of devices.  

\begin{figure}[htp]
\centerline{\includegraphics[width=8cm]{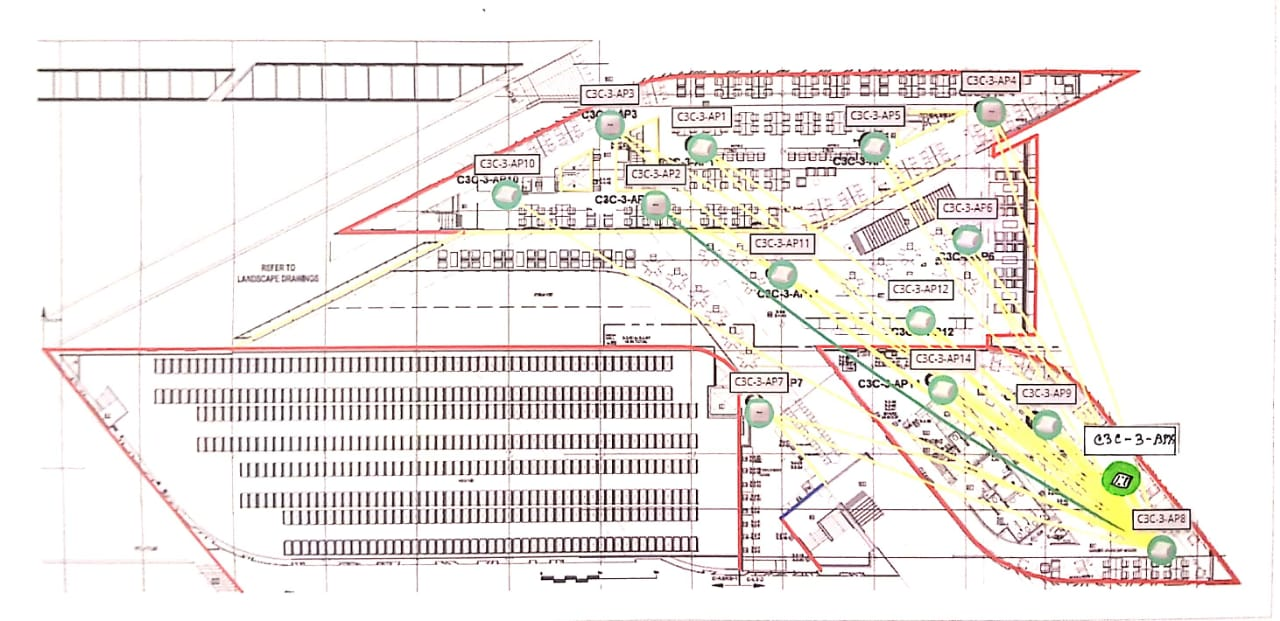}}
\caption{Floor plan with proposed new Wi-Fi access point in library L-3}
\label{fig11}
\end{figure}

Next, after spotting the data trend for library L-4, we could present a proposal to place a new access point near the inspected access point (C3C-4-AP5). The new floor plan with the new AP marked as C3C-4-APX is shown in figure 12. To activate this new access point in the proposed location we will require approximately eight (08) meters of Ethernet cable to cover the distance between the new access point and the nearest communications room (room no. 496) and the newly installed access point will be serving a rational number of Wi-Fi users and certainly reduce the excess pressure on C3C-4-AP5. 

\begin{figure}[htp]
\centerline{\includegraphics[width=8cm]{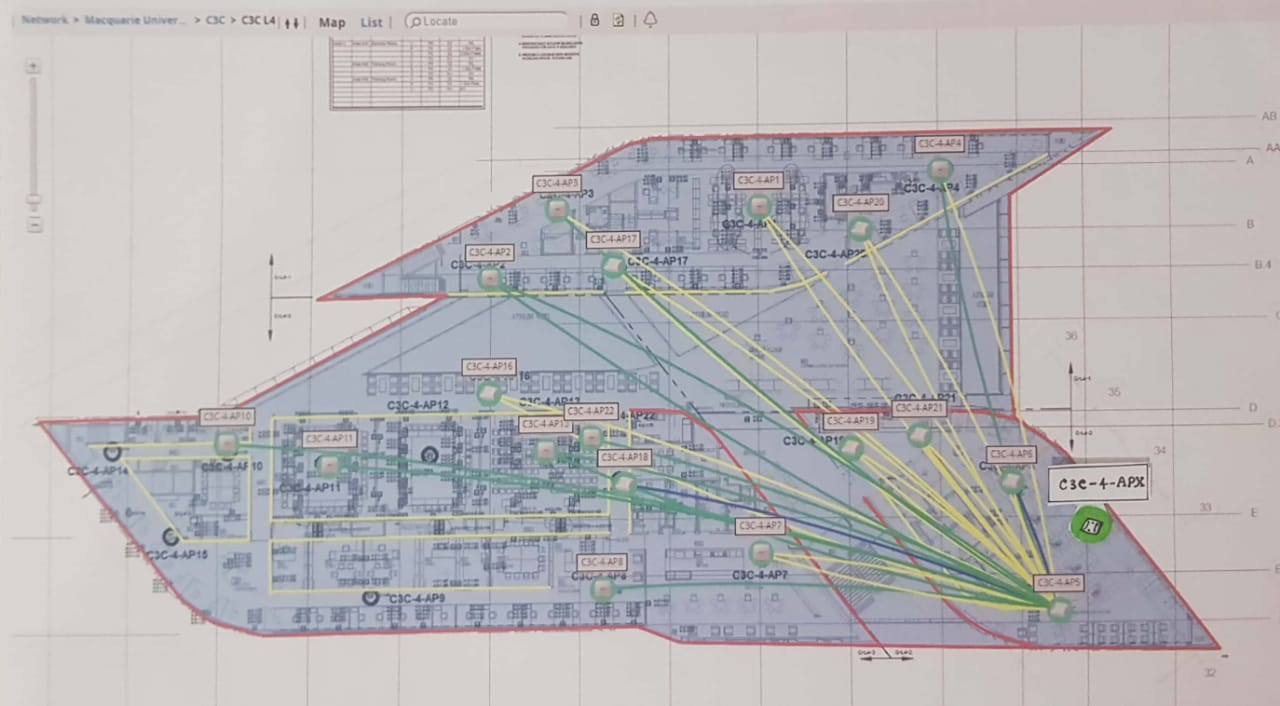}}
\caption{Floor plan with proposed new Wi-Fi access point in library L-4}
\label{fig12}
\end{figure}

\section{Conclusion and Future Work}

This paper widely focuses on the wireless measurements which is used as a measure for finding out the locations in the university campus where the current Wi-Fi network requires improvements. Using few wireless measurements tools, several coverage maps were generated in various locations showing the signal strength levels of the same. Next, a range of data related to the access points were collected from the Management Platform and analyzed accordingly. These stages contribute towards identifying the over-crowded access points as well as the locations with weak or no Wi-Fi signal coverage. Finally, few plans for installing new access points in some widely used locations around the campus has been presented along with the coverage prediction of improved signal quality. 

Apart from obvious results obtained throughout the study, we like to bring up a comparative discussion on the theoretical specifications and their practical implementations:

- According to a campus wireless networks planning guideline, in a high-density (HD) campus environment, ideally there should be between 40 to 60 users for each radio per access point \cite{b11}. But if we look at the practical AP deployments, the number of clients associated to any access point hardly matches with these given numbers. Numerous accounts could be identified to explain this situation; firstly, the maximum number of clients that can be served by a single access point highly depends on the fact that what internet services are being availed by the users. Secondly, sometimes users can associate with an access point located in a distance area instead of connecting to a nearby access point which is offering weak signal strength due to co-channel interference or interference caused by walls or other solid substances. Next, another possible scenario in a high-density area, for example, an auditorium or a library is, users are not equally distributed over the area. Hence, any one access point can experience congestion while keeping another access point idle. Finally, the coverage range and the maximum number of clients that can connect to a specific access point, depends on the model of the access point to a great extent. 

- Theoretically to make the high density (HD) campus deployment Voice and Roaming optimized, it is ideal to place access points in a honeycomb structure, with a distance of 50 feet between any two of them to guarantee great density of access points for users to travel without disturbing continuous application execution \cite{b11}. But practically its clearly seen that this idea is not utilized while placing the access points (figure 11 and figure 12). The reasons behind this disparity between theoretical and real-world aspects can be perceived from various aspects, for example; it becomes quite difficult to establish a complete hypothetical structure for the wireless network before the construction of the buildings takes place and yet again it cannot be always matched with the architectural plan of the buildings such as position of the walls and dividers, placing of furniture and fixtures, heights of the ceilings, widths of corridors etc. Again, most of the buildings around the campus have been constructed years ago, even before the Wi-Fi came into being, hence these building structures can barely be changed according to the wireless network requirements. 

It can be concluded that with implementation of the proposed plans for new access points deployments, it is expected that the campus wireless network will become more stronger and be able to provide the staffs, students and guests with the best wireless experience at any location around the campus. Again, the existing wireless network system will keep requiring modifications that can be made effortlessly to cope with the ever-developing Wi-Fi technology and its newly-added versions.

\end{document}